\documentclass[journal=estlcu,manuscript=article]{achemso}

\usepackage[version=4]{mhchem} % Formula subscripts using \ce{}
\usepackage{color}
\usepackage{makecell}
\usepackage{hyperref}
\usepackage{cleveref}
\usepackage{amsmath}
\usepackage{float}
\usepackage{pdfpages}
\title[diffmix]
  {Differentiable Modeling and Optimization of Battery Electrolyte Mixtures Using Geometric Deep Learning}

\author{Shang Zhu}
\affiliation{Department of Mechanical Engineering,\\
 Carnegie Mellon University, Pittsburgh, Pennsylvania 15213, USA}
\author{Bharath Ramsundar}
\affiliation{Deep Forest Sciences, Palo Alto, California 94306, USA}
\author{Emil Annevelink}
\affiliation{Department of Mechanical Engineering,\\
 Carnegie Mellon University, Pittsburgh, Pennsylvania 15213, USA}
\author{Hongyi Lin}
\affiliation{Department of Mechanical Engineering,\\
 Carnegie Mellon University, Pittsburgh, Pennsylvania 15213, USA}
\author{Adarsh Dave}
\affiliation{Department of Mechanical Engineering,\\
 Carnegie Mellon University, Pittsburgh, Pennsylvania 15213, USA}
\author{Pin-Wen Guan\footnote{Current affiliation: Sandia National Laboratories, Livermore, California 94550, USA}\ }
\affiliation{Department of Mechanical Engineering,\\
 Carnegie Mellon University, Pittsburgh, Pennsylvania 15213, USA}

\author{Kevin Gering}
\affiliation{Energy Storage \& Technology, Idaho National Laboratory, Idaho Falls, Idaho 83415, USA}
\author{Venkatasubramanian Viswanathan}
\email{venkvis@cmu.edu}
\affiliation{Department of Mechanical Engineering,\\
 Carnegie Mellon University, Pittsburgh, Pennsylvania 15213, USA}

\begin{document}
%%%%%%%%%%%%%%%%%%%%%%%%%%%%%%%%%%%%%%%%%%%%%%%%%%%%%%%%%%%%%%%%%%%%%
%% The abstract environment will automatically gobble the contents
%% if an abstract is not used by the target journal.
%%%%%%%%%%%%%%%%%%%%%%%%%%%%%%%%%%%%%%%%%%%%%%%%%%%%%%%%%%%%%%%%%%%%%
\begin{abstract}
Electrolytes play a critical role in designing next-generation battery systems, by allowing efficient ion transfer, preventing charge transfer, and stabilizing electrode-electrolyte interfaces.
In this work, we develop a differentiable geometric deep learning (GDL) model for chemical mixtures, \textbf{DiffMix}, which is applied in guiding robotic experimentation and optimization towards fast-charging battery electrolytes. In particular, we extend mixture thermodynamic and transport laws by creating GDL-learnable physical coefficients. We evaluate our model with mixture thermodynamics and ion transport properties, where we show improved prediction accuracy and model robustness of \textbf{DiffMix} than its purely data-driven variants. Furthermore, with a robotic experimentation setup, \textbf{Clio}, we improve ionic conductivity of electrolytes by over 18.8\% within 10 experimental steps, via differentiable optimization built on \textbf{DiffMix} gradients. By combining GDL, mixture physics laws, and robotic experimentation, \textbf{DiffMix} expands the predictive modeling methods for chemical mixtures and enables efficient optimization in large chemical spaces.
\end{abstract}

% \keywords{Chemical Mixtures $|$ Differentiable Physics $|$  Geometric Deep Learning $|$ Gradient-based Optimization} 
%%%%%%%%%%%%%%%%%%%%%%%%%%%%%%%%%%%%%%%%%%%%%%%%%%%%%%%%%%%%%%%%%%%%%
%% Start the main part of the manuscript here.
%%%%%%%%%%%%%%%%%%%%%%%%%%%%%%%%%%%%%%%%%%%%%%%%%%%%%%%%%%%%%%%%%%%%%

% \section*{Introduction}
Chemical mixtures are widely used in chemical processes and devices such as energy storage and conversion\cite{le1,le2,osc1,osc2}, chemical reactions and catalysis\cite{react_textbook, react_exp, catalysis1}, and environmental engineering\cite{es1,es2,es3}. Often, the mixture chemistry and compositions are carefully designed to achieve higher device performances. In particular, battery electrolytes, as mixtures of salts and solvents, have been optimized to facilitate ion transport, prevent electron transfer, and stabilize electrode-electrolyte interfaces for an energy-dense and durable battery system\cite{le3,le4, electrolyte_review_science,automat}. %Specifically, electrolyte ionic conductivities determine the fast charging capabilities of batteries and researchers have optimized the electrolyte chemistry to increase the ionic conductivities, especially in low temperature operations\cite{chunsheng_nature}.

The design and optimization of electrolyte mixtures remain challenging due to the complexity of mixture chemistry and compositions, as well as the high experimentation cost\cite{otto, clio}. Physics-based modeling offers a solution by probing the underlying mixture physics and rationalizing the design principles for high-performing mixtures. Among physics-based mixture modeling techniques, molecular simulation is a powerful tool to study the interactions and dynamic evolution inside a complex mixture system, but it can be limited to time and length scales due to its high computational costs\cite{md1, yumin_md}. Alternatively, chemical physicists proposed empirical function relationships to describe mixture physics. For example, Redlich-Kister (R-K) polynomials\cite{r-k} were designed for modeling mixture thermodynamics, and Arrhenius equation\cite{arr} was proposed to describe the temperature dependence of chemical reactions and other dynamic behaviors. Although they may provide decent model accuracy and indicate intrinsic physical behaviors such as reaction energy barriers, these empirical relationships are lacking in predictive power when new chemical species are provided.
% by considering the composition symmetries and provided an extension to higher order polynomials to capture the complex behaviors of mixture properties. Empirical evidence showed that the fitted coefficients in front of the polynomials may indicate the mixture interaction type (non-associating, associating, inter-component associating)\cite{r-k}. This approach relies on fitting experimental data to obtain the physical coefficients\cite{emv1,emv2,ee1,ee2}, which may not be applicable when evaluated on an unexplored chemical space. More rigorously, one can derive chemical physics laws based on atomic microstates by molecular-level simulation and correlate them with macroscopic mixture properties\cite{pure_stat,mix_stat,fp2, fp1}. 
Emerging data-driven methods\cite{qspr,Yao2023,Pablo-Garcia2023,Levin2022,Goldman2023,Chen2021,Chen2022} can potentially bridge the gap in the predictive modeling of electrolyte mixtures \cite{ml_vis, hb1, hb2, rafael2}. Notably, with a linear regression method, Kim et al.\cite{dd_electrolytes_pnas} discovered a strong correlation between the oxygen content in battery electrolytes and lithium-metal-cell Coulombic efficiencies. Bradford et al.\cite{polye_acs} developed a graph machine learning model of solid polymer electrolytes (SPEs) and predicted ionic conductivities of thousands of new SPEs. 
% and improve the training accuracy % In terms of chemical mixtures, permutation invariance must be preserved over the components and compositions.%In molecular machine learning, geometric priors through the framework of geometric deep learning (GDL) simultaneously constrain the model space and enable flexible embeddings of the atomic environment\cite{gdl1,gdl2}. GDL methods have demonstrated state-of-the-art model performances when predicting benchmark materials and molecular properties\cite{mpnn, cgcnn, m3gnet}. 
%, thanks to the availability of large scale scientific datasets\cite{molnet,mp}. 
% In terms of chemical mixtures, Molecular mixtures are less explored than pure molecule and material substances\cite{cgcnn,mpnn,molnet,megnet} due to the challenges of handling high-dimensional chemical and compositional space. 
Furthermore, the differentiability of modern deep learning models provides a new opportunity for unifying physics-based and data-driven models \cite{jaxmd2020,MANN2022108232,PhysRevLett.127.126403,GUAN2022114217,10.1063/5.0126475,Shen2023}. Especially, Guan proposed a general differentiable framework merging thermodynamic modeling and deep learning for multi-component mixtures, where all the thermodynamic observables including thermochemical quantities and phase equilibria can be auto-differentiated, thus allowing models learned by gradient-based optimization\cite{GUAN2022114217}. It was subsequently extended to a more comprehensive framework of differentiable materials modeling and design, %which seamlessly integrates physics-based computation and deep learning, and includes
including the full processing-structure-properties-performance relationships\cite{diffmat}.%\pg{It was subsequently extended to a more comprehensive framework of differentiable materials modeling and design, including the full chain of processing-structure-property\cite{diffmat}.} %The framework was demonstrated in the Cu-Rh system and the mixing parameters were successfully learned\cite{GUAN2022114217}.} 

In this work, we leverage the geometric deep learning (GDL) method for battery electrolyte modeling and optimization, % to create the map from chemical species to physics laws, 
where, in GDL, necessary geometric priors are applied as constraints on the model space to improve model efficiency\cite{gdl1,gdl2}. In particular, we develop a differentiable GDL model of chemical mixtures, \textbf{DiffMix}, which is applied in guiding the robotic experimentation towards fast-charging battery electrolytes. %by combining GDL and mixture physics laws, 
The GDL component is designed to transform the molecular species, compositions, and environment conditions, to physical coefficients in predefined mixture physics laws, where the Redlich-Kister (R-K) mixing theory and Vogel–Fulcher–Tammann (VFT) model are selected for mixture thermodynamic and transport properties, respectively. % Then we develop graph neural networks (GNNs) to encode molecular species and compositions and predict the physical coefficients inside the selected mixture physics model, with the component-wise permutation invariance preserved. 
We test the predictive power of \textbf{DiffMix} on a non-electrolyte binary mixture dataset of excess molar enthalpies and excess molar volumes, and thereafter on a large-scale simulation dataset of electrolyte ionic conductivities.
% multicomponent battery electrolyte mixtures. 
We compare our model with its purely data-driven methods and show superior performances on prediction accuracy and robustness. 
Further, with our previously built robotic experimentation setup, \textbf{Clio}\cite{clio}, we demonstrate a differentiable optimization on battery electrolyte mixtures, based on the gradient information from \textbf{DiffMix} auto-differentiation. We successfully improve the ionic conductivity values by over 18.8\% within 10 experimental steps in the evaluated chemical space, enabling the fast-charging design of battery systems. Our method extends the modeling techniques of battery electrolyte mixtures by unifying physics models and geometric deep learning and, to the best of our knowledge, realizing the first differentiable optimization of battery electrolyte properties. %Lastly, the simulated optimization runs are validated by o indicating the robustness of our differentiable mixture physics framework.

\subsection*{DiffMix: Combining Physics and Geometric Deep Learning for Modeling Chemical Mixtures}
\begin{figure*}[htb!]
\centering
\includegraphics[width=\linewidth]{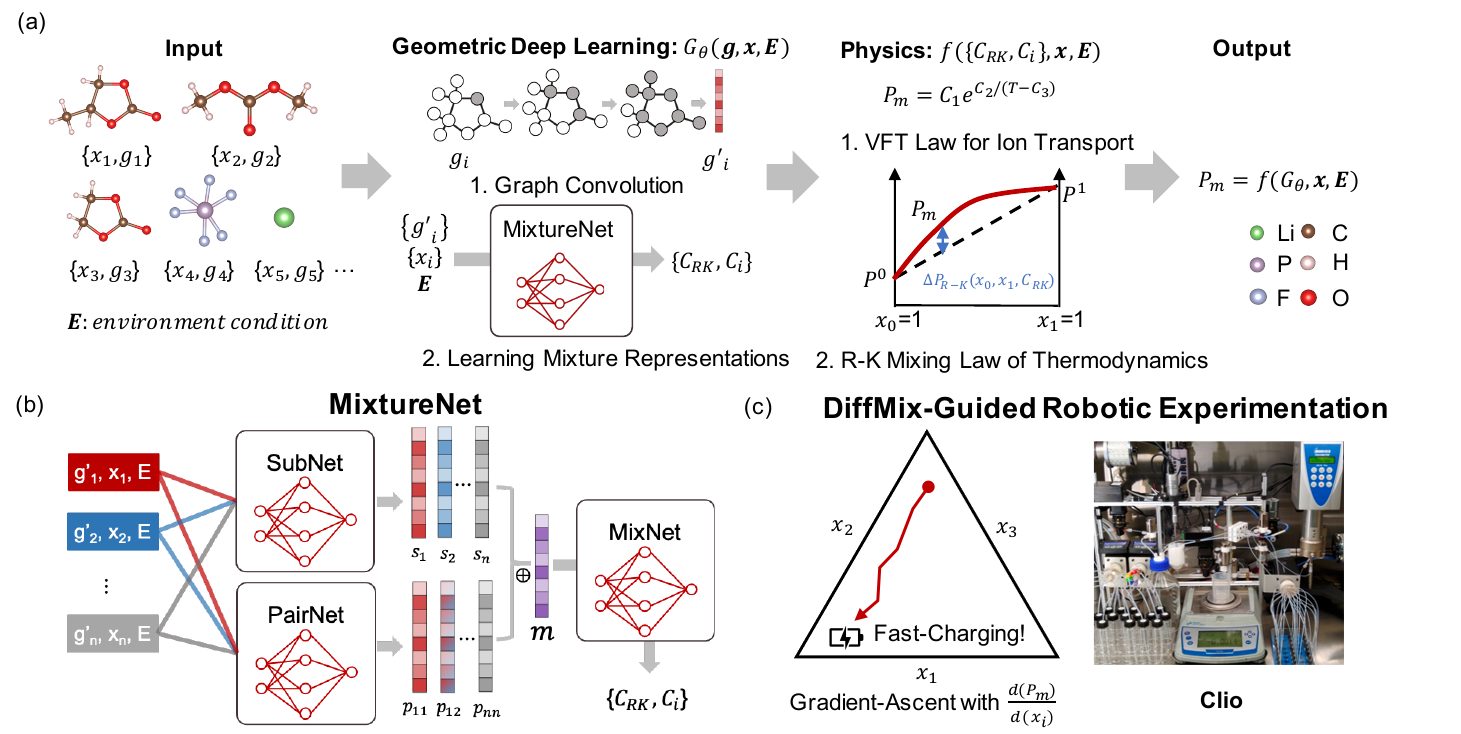}% Here is how to import EPS art
\caption{\label{F1}Differentiable Modeling and Optimization of Chemical Mixtures with \textbf{DiffMix}. (a)Model Architecture. Input: chemical graphs $\mathbf{g}=\{g_i\}$, compositions $\mathbf{x}=\{x_i\}$ and environment condition vector $\mathbf{E}$ (e.g. temperature, pressure). Output: mixture property $P_m$. \textbf{DiffMix} combines geometric deep learning and physics laws in a differentiable framework. The GDL component, $G_\theta(\cdot)$, transforms $(\mathbf{g}, \mathbf{x}, \mathbf{E})$ to coefficients in physics laws, $\{C_{RK},C_i\}$, via graph convolutional operations on graphs and \textbf{MixtureNet} that convolves both chemical identities and $(\mathbf{x}, \mathbf{E})$ to learn mixture representations. $\theta$ is the set of learnable parameters. 
Two example physics laws, $f(\{C_{RK},C_i\}, \mathbf{x}, \mathbf{E})$, VFT law and R-K mixing law (binary as an example), are included here but can be generalized. Overall, the mixture property output can be written as $P_m=f(\mathbf{G_{\theta}}, \mathbf{x}, \mathbf{E})$. (b) Detailed architecture of \textbf{MixtureNet}. Input is $(\mathbf{g'}, \mathbf{x}, \mathbf{E})$, where $\mathbf{g'}$ is the graph embedding after graph convolutions. Input is processed by weight-sharing fully connected neural networks (FCNN), \textbf{SubNet} and \textbf{PairNet}, to learn the per-substance and pairwise-interaction embeddings, $\{s_i\}$ and $\{p_{ij}\}$, respectively. Mixture embeddings $m$ is created after a pooling operator ($\oplus$) on $\{s_i, p_{ij}\}$, and followed by another FCNN, \textbf{MixNet}, to produce physical coefficients. The design of  $\oplus$ and \textbf{MixNet} depends on the downstream physics laws. (c) Differentiable Optimization and Robotic Experimentation for Fast-charging Battery Electrolytes. With a trained \textbf{DiffMix} on battery electrolyte ionic conductivities, auto-differentiation provides the gradient information of $\frac{d(P_m)}{d(x_i)}$ over input compositions. We run a gradient-ascent algorithm on composition space and guide a robotic experimentation setup, \textbf{Clio}\cite{clio}, for fast-charging battery electrolyte design.} % Taking an electrolyte system of one salt and ternary co-solvents as an example
\end{figure*}

Our model, \textbf{DiffMix}, combines physics and geometric deep learning in order to build a differentiable and predictive model for chemical mixtures, as shown in \Cref{F1} (a). Taking the input of chemical graphs, compositions, and environment condition vector $(\mathbf{g}, \mathbf{x}, \mathbf{E})$, \textbf{DiffMix} processes with two components, geometric deep learning $G_\theta(\mathbf{g}, \mathbf{x}, \mathbf{E})$ and physics laws, $f(\cdot,\mathbf{x}, \mathbf{E})$, and then output mixture property $P_m=f(\mathbf{G_{\theta}}, \mathbf{x}, \mathbf{E})$, in an end-to-end differentiable framework.

\textbf{Physics Models for Thermodynamics of Mixing and Ion Transport}. The selection of physics models, $f(\cdot,\mathbf{x}, \mathbf{E})$, depends on the mixture properties of interest. Here, we take the mixing thermodynamics of binary non-electrolyte mixtures and ion transport of multicomponent electrolyte mixtures as examples, which can be further generalized to other forms\cite{Thomas1984,SIEGEL2021807}. 

%\subsection*{Polynomial-based Thermodynamic Laws for Chemical Mixtures} 
To describe the thermodynamics of mixing of non-electrolyte mixtures, %we construct a lattice-based model system of mixtures, where the thermodynamic quantity change is thus driven by the interactions between lattices. Taking the inspiration of thermodynamics of mixing in binary mixtures, e.g. enthalpy change during mixing, we can obtain that the excess molar property, $\Delta P^m$, as a simplified function of mole fractions of each component.
% \begin{align}
% \Delta P^m=R^0_{12}\cdot x_1x_2 \label{eq1}
% \end{align}
% where $x_1$, $x_2$ are mole fractions of species 1 and 2, and $R^0_{12}$ is the interaction factor that depends on the lattice volume, interaction energy coefficient, coordination number, etc. A more detailed derivation for excess molar enthalpy can be found in Supporting Information (SI).
a polynomial expansion can be used for representing the excess function of mixing $\Delta P_m$, i.e. the difference between mixing thermodynamic quantity $P_m$ and the linear combination of each component $\Sigma_{i}x_iP^i$, where $P^i$ is the property of the species $i$. It has been successfully applied in differentiable thermodynamic modeling\cite{GUAN2022114217}, with the Redlich-Kister (R-K) polynomial\cite{r-k} being a popular choice:

\begin{align}
{\Delta P_m}= \Sigma_{i<j} [x_ix_j\Sigma^N_{k=0}C^k_{RK,ij}(x_i-x_j)^k] \label{eq2}
\end{align}
where $x_i$ and $x_j$ are mole fractions of species $i$ and $j$, $C^k_{RK,ij}$ is the R-K polynomial coefficients between the two species and with order number $k$. \Cref{eq2} preserves the permutation invariance of chemical species i and j, when the odd orders of polynomials follow the parity rule of permutation. %Empirical evidence show that $C^k_{R-K,ij}$ may indicate the mixture interaction strengths\cite{r-k}, but they are lacking in predictive power when exploring a new chemical space.
The mixture thermodynamic property $P^m$ can be further obtained by:
\begin{align}
P_m=\Sigma_{i<j}[x_ix_j\Sigma^N_{k=0}C^k_{RK,ij}(x_i-x_j)^k]+\Sigma_{i}x_iP^i\label{eq3}
\end{align}
\Cref{eq3} preserves permutation invariance over mixture components and can be applied to a wide range of mixture thermodynamic properties.
% \subsection*{Vogel–Fulcher–Tammann (VFT) Model for Mixture Transport Properties}
%Transport phenomena in chemical mixtures, as opposed to thermodynamic properties, depend on the reorganization of microstates through the transition state\cite{yumin_md}. 

On the ion transport properties, we focus on the ionic conductivities of battery electrolytes. A higher ionic conductivity will reduce the ion transfer resistance between electrodes and lessen the formation of electrolyte concentration polarization, therefore enabling fast-charging battery applications\cite{chunsheng_nature}. Here, we select the Vogel–Fulcher–Tammann (VFT) model to capture the temperature dependence \cite{VFT} as:
\begin{align}
P_m =C_1 e^{-\frac{C_2}{T-C_3}} \label{eq4}
\end{align}
where $T$ is the temperature and $\{C_i\}$ is a set of physical coefficients.

\textbf{Geometric Deep Learning to Learn Mixture Representations}. \Cref{eq3,eq4} describe thermodynamic and ion transport laws that conventionally rely on empirically fitting experimental data to obtain physical coefficients, $\{C_{RK},C_i\}$. However, the function relationship between mixture input $(\mathbf{g}, \mathbf{x}, \mathbf{E})$ and physical coefficients $\{C_{RK},C_i\}$ remains unknown. GDL component is therefore introduced to replace physical coefficients with learnable GDL functions, $\{C_{RK},C_i\}=G_\theta(\mathbf{g}, \mathbf{x}, \mathbf{E})$. In this way, the mixture physics model now becomes predictive and fully differentiable from chemical structures to properties. The first step in the GDL component is a graph convolution \cite{graphconv} transformation over each component graph $g_i$ to obtain the graph-level feature vector $g'_i$, for component $i$. In the second step, $g'_i$ is attached with compositions and environment conditions and processed by \textbf{MixtureNet} to learn the mixture-level representations. \textbf{MixtureNet} architecture is shown in \Cref{F1} (b). Each attached mixture component vector $[g'_i,x_i,\mathbf{E}]$ passes through two weight-sharing fully connected neural networks (FCNN), \textbf{SubNet} and \textbf{PairNet}, to learn the per-substance and pairwise-interaction embeddings, $\{s_i\}$ and $\{p_{ij}\}$, respectively. Depending on the mixture physics laws, $\{s_i\}$ and $\{p_{ij}\}$ are combined in a certain form to produce the physical coefficients. For VFT model in \Cref{eq4} and battery electrolyte mixtures, mixture feature vector $m$ is created via a pooling operator $\oplus$, $m=\{s_i\}\oplus\{p_{ij}\}=[\Sigma_ix_i\cdot s_i,\Sigma_{ij}x_ix_j\cdot p_{ij}]$, by concatenating the weighted sums of substance and pair embeddings. The physical coefficients in the VFT model, $\{C_i\}$, is a function of $m$ via another FCNN, \textbf{MixNet}. For the mixing law of thermodynamics and R-K polynomial-based model in \Cref{eq3}, due to the intrinsic per-substance dependence of $P^i$ and pairwise interaction dependence of $\{C^k_{RK,ij}\}$, they can be produced directly from \textbf{SubNet} and \textbf{PairNet} without additional pooling operations or \textbf{MixNet}. More details about the model implementation can be found in the Methods section. In the GDL component, we preserve permutation invariance over components with the mixture pooling operator ($\oplus$) in the VFT model and the intrinsic permutation invariance introduced in the R-K model.

To benchmark the effectiveness of combining physics laws with data-driven models, we design a purely data-driven baseline, \textbf{GNN-only}, by removing the mixture physics model in \textbf{DiffMix}. In the VFT-type of the GDL component, instead of outputting $\{C_i\}$ as the physics law coefficients, the \textbf{GNN-only} model ignores the physics laws and directly produces the mixture properties. All models are evaluated on two thermodynamic datasets of binary non-electrolyte mixtures and one transport property dataset of battery electrolyte mixtures. The thermodynamic data include literature-curated excess molar enthalpies (631 data points) and excess molar volumes (1,069 data points). For electrolytes, the ionic conductivity dataset is prepared that contains 24,822 mixtures of single-salt-ternary-solvent electrolyte solutions, generated by the Advanced Electrolyte Model (AEM)\cite{aem,GERING20063125}. More data generation details can be found in the Methods part. In Supplementary Information (SI), we further test a data-driven variant with Morgan fingerprints for molecules\cite{ecfp}.

\textbf{DiffMix-Guided Robotic Experimentation and Optimization for Battery Electrolytes}.
Differentiability enables gradient-based optimization for materials modeling and design\cite{GUAN2022114217,diffmat}. With auto-differentiation on a trained \textbf{DiffMix} model, we can conveniently obtain gradient information of mixture property output over input compositions, $\frac{d(P_m)}{d(x_i)}$, and thereafter navigate the mixture chemical space in order to optimize the mixture property objective. In \Cref{F1} (c), we demonstrate the battery electrolyte optimization on a ternary co-solvent composition space to maximize the ionic conductivity via a gradient-ascent algorithm, and guide our previously developed robotic experimentation setup, \textbf{Clio}, to improve the electrolyte ion transport properties for fast-charging batteries.

% \section*{Results}

\subsection*{Differentiable Modeling on Thermodynamic and Transport Properties of Chemical Mixtures} 
\begin{figure}[htb!]
\centering
\includegraphics[width=\linewidth]{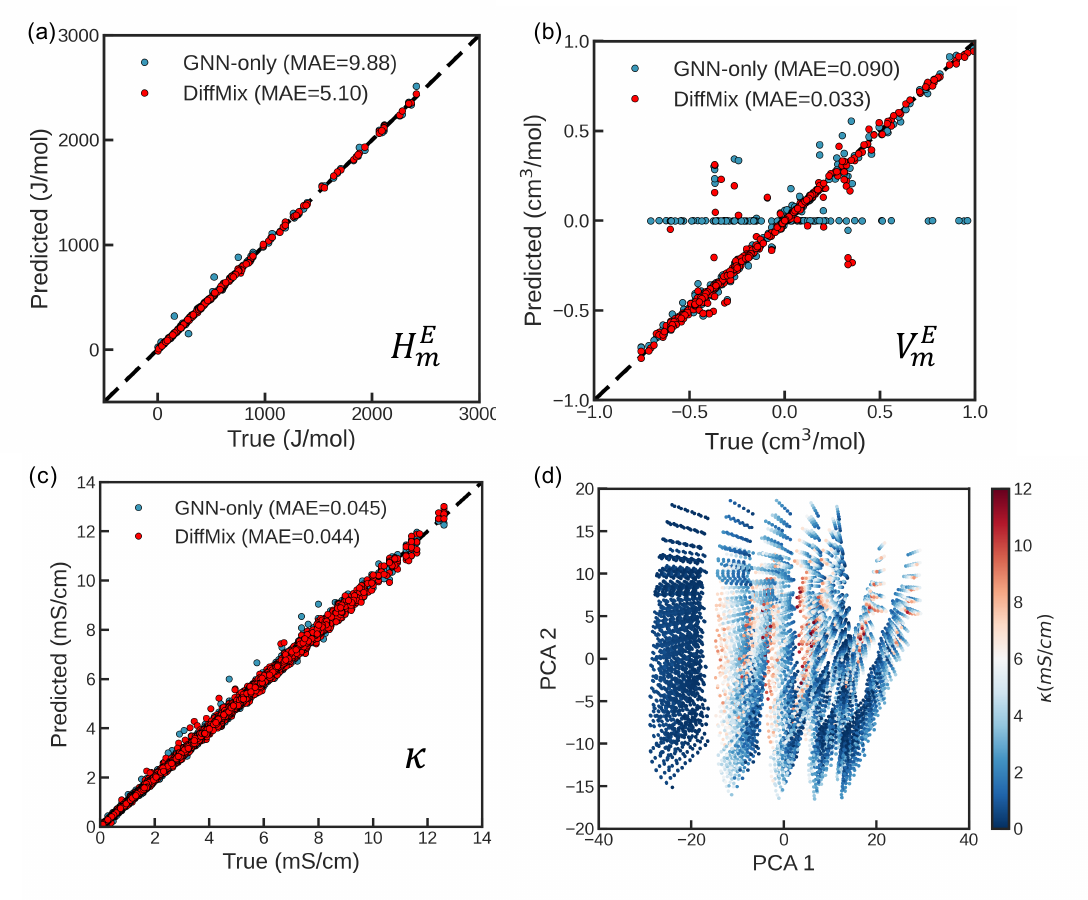}
\caption{\label{F2} Prediction Accuracy Analysis of \textbf{DiffMix}. Parity Plots of (a) Excess Molar Enthalpy ($H_m^E$) Testing Dataset, (b) Excess Molar Volume ($V_m^E$) Testing Dataset, and (c) Ionic Conductivity ($\kappa$) Testing Dataset. (d) Two-dimensional Principal-Component-Analysis (PCA) of Mixture Features Extracted from the Trained \textbf{DiffMix} Model on the Full Ionic Conductivity Dataset.}
\end{figure}

We start our result analysis on excess molar enthalpies ($H_m^E$) and excess molar volumes ($V_m^E$) of binary non-electrolyte mixtures. The model performances of \textbf{DiffMix} and \textbf{GNN-only} model are summarized in \Cref{tbl:thermo}. We confirm the permutation invariance of both models, considering the identical loss values before and after permuting the component sequences.
Further, we find that \textbf{DiffMix}, built on the known physics prior, outperforms the \textbf{GNN-only} model by a noticeable margin. We achieve mean-absolute-errors (MAEs) of $0.033\pm0.009$ (cm$^3$/mol) and $5.10\pm0.32$ (J/mol) for excess molar volumes and excess molar enthalpies, respectively, with their parity plots shown in \Cref{F2} (a-b). % %Therefore, with \textbf{DiffMix}, we already approach the accuracy limit of excess molar enthalpy prediction.

\begin{table}
  \centering
  \caption{Model Performance on Thermodynamic Properties and Ionic Conductivities\textsuperscript{\emph{a}}}
  \label{tbl:thermo}
  \begin{tabular}{ccccccc}
    \hline
    \hline
    Task & \makecell{DiffMix\\test} & \makecell{DiffMix\\permuted\textsuperscript{\emph{c}}}  & \makecell{GNN-only\\test} & \makecell{GNN-only\\permuted\textsuperscript{\emph{c}}}\\ 
    \hline
    \makecell{$V_m^E$\\(cm$^3$/mol)} &  \makecell{\textbf{0.033}\\\textbf{$\pm$0.009}\textsuperscript{\emph{b}}} &  \makecell{\textbf{0.033}\\\textbf{$\pm$0.009}\textsuperscript{\emph{b}}}   & \makecell{0.090\\$\pm$0.106\textsuperscript{\emph{b}}} & \makecell{0.090\\$\pm$0.106\textsuperscript{\emph{b}}} \\
    \hline
    \makecell{$H_m^E$\\(J/mol)} &  \makecell{\textbf{5.10}\\\textbf{$\pm$0.32}\textsuperscript{\emph{b}}}  & \makecell{\textbf{5.10}\\\textbf{$\pm$0.32}\textsuperscript{\emph{b}}} & \makecell{9.88\\$\pm$2.21} & \makecell{9.88\\$\pm$2.21} \\
    \hline
    \makecell{$\kappa$\\(mS/cm)} &  \makecell{\textbf{0.044}\\\textbf{$\pm$0.005}}&  \makecell{\textbf{0.044}\\\textbf{$\pm$0.005}} & \makecell{0.045\\$\pm$0.006}& \makecell{0.045\\$\pm$0.006} \\
    \hline
    \hline
  \end{tabular}
  
  \raggedright{$a$. Results are reported by regression mean-absolute-errors (MAEs) (Mean $\pm$ Standard Deviation) after running an ensemble of 5 models. $b$. The polynomial order $N$ is 4 in \Cref{eq3}. $c$. Permuted loss is generated by permuting input component sequences.}
\end{table}

Further, we investigate the predictive power of \textbf{DiffMix} on ionic conductivities ($\kappa$) of multi-component electrolyte solutions. With the 24,822 ionic conductivity data points, we train \textbf{DiffMix} and compare it with the \textbf{GNN-only} baseline model. The prediction accuracy on the testing sets is shown in \Cref{tbl:thermo}, where we can further confirm the permutation invariance in \textbf{DiffMix}. %In \textbf{MixEFCP}, by permuting the sequences of solvents, the testing MAE jumps from $0.105 \pm 0.007$ (mS/cm) to $ 22.70 \pm 6.62$ (mS/cm). 
With the physics-incorporated \textbf{DiffMix} model, we achieve the lowest MAE, $0.044$ (mS/cm) considering the maximum ionic conductivity above 12 (mS/cm) in the training set, as shown in the parity plot of \Cref{F2} (c). Compared with thermodynamic results, the accuracy improvement by adding physics priors is not as significant here. This may be attributed to the limited physical capacity of the VFT model in \Cref{eq4}, but further investigation is required, such as testing alternative physics laws for ionic conductivities. Lastly, in \Cref{F2} (d), we visualize the learned mixture features ($m$ in \Cref{F1}b) for ionic conductivities with principal component analysis (PCA) in two dimensions. We observe a smooth distribution of high and low $\kappa$ values, indicating a good discriminative power of the trained \textbf{DiffMix} model.

\subsection*{Physics Model Capacity and Temperature Extrapolation} 

\begin{figure}[!p]
\centering
\includegraphics[width=\linewidth]{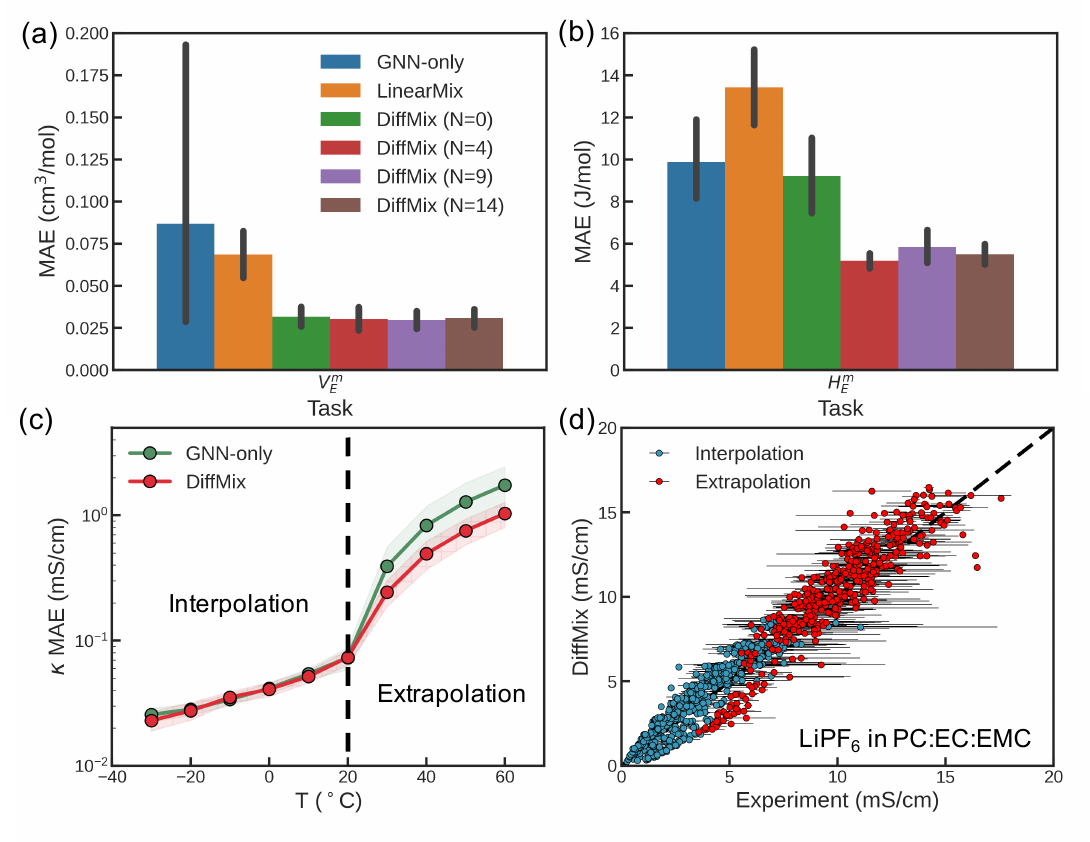} %Modeling Transport Properties of Electrolyte Mixtures:
\caption{\label{F3} (a-b) Physics Law Analysis on Thermodynamic Data. Varying the polynomial order for (a) excess molar volume ($V_m^E$), and (b) excess molar enthalpy ($H_m^E$). In both cases, the regression MAEs of \textbf{DiffMix} models with N=$0, 4, 9, 14$ are compared with those of \textbf{GNN-only} and linear mixing models. The bar plots are generated after running an ensemble of 5 models, where the black lines display the standard deviations of results. (c-d) Model Extrapolation on Ionic Conductivity Data. (c) Prediction Accuracy of \textbf{DiffMix} and \textbf{GNN-only} for interpolation and extrapolation cases. Regression MAEs are grouped by data points with the same temperatures. The training is performed on low-temperature AEM data ($\leq20\ ^\circ$C) so the evaluations at higher temperatures are viewed as extrapolation. (d) Parity plots of \textbf{DiffMix} predictions with experimental measurements for electrolyte solutions where \ce{LiPF6} salts are dissolved in PC:EC:EMC ternary solutions\cite{exp_ion}. Blue dots are interpolation cases, while red dots are extrapolation cases. Black solid lines are standard deviations of the reported experimental values. }
\end{figure}

For the mixture thermodynamics tasks, so far, the polynomial order $N$ in \Cref{eq3} is specified as four. %, while the lattice-based theory in \Cref{eq1} indicates that $N=0$ may be enough. 
To study the polynomial-order dependence of model capacity, we vary the polynomial order as $N=0, 4,9,14$ or fully remove the excess term in \Cref{eq3}.  The latter essentially describes the linear mixing rule. The results are shown in \Cref{F3} (a) and (b), where we also compare them with \textbf{GNN-only} in order to see the effectiveness of the added physics models.  First, we observe the trend of decreasing testing errors when higher orders of polynomials are introduced, i.e. increasing the capacity of the mixture physics model. With $N=4$, MAEs for both $V_m^E$ and $H_m^E$ get reduced by over half than those of the linear mixing model. However, the model performance plateaus as we further increase the polynomial-based model capacity. It is worth noting that the experimentation uncertainty is around 0.005 (cm$^3$/mol) and 5 (J/mol) for the two measurements. For the excess molar volume task, the plateauing behavior may be due to the fact that \textbf{DiffMix} accuracy is limited by the GDL model capacity. However, for the enthalpy task, it can also be attributed to the data uncertainty, considering that the \textbf{DiffMix} prediction MAE is close to the measurement error. Compared with the \textbf{GNN-only} baseline model, even the linear mixing model displays a lower MAE in the excess molar volume task, while in the enthalpy case, adding the zeroth-order interaction terms improves the worst performing linear mixing model so it outperforms the \textbf{GNN-only} baseline. This may be explained by that the enthalpy change relies on the inter- and intra- molecular interactions between lattices, modeled by the pair-wise interaction coefficients in R-K mixing laws, but the mixing volume property more relies on the property of individual components. Further, Figure S 2 describes the overall decreasing trend of R-K polynomial coefficients $\{C_{RK,ij}^k\}$ when 15 polynomials are included in the physics-based R-K model, explaining the plateauing pattern of the model accuracy.

On modeling the ionic conductivities ($\kappa$) of battery electrolytes, we test model extrapolation to higher temperatures, as shown in \Cref{F3} (c). In \Cref{F3} (c), we report the prediction MAEs grouped by temperatures in the range of [-30 $^\circ$C, 60 $^\circ$C]. Note that our training is performed on the data with temperature range [-30 $^\circ$C, 20 $^\circ$C]. We notice that the interpolation MAE is close to 0 for both models, consistent with the low MAE results reported in \Cref{tbl:thermo}. However, in the extrapolation test on the data generated above 20 $^\circ$C, non-negligible errors have been detected, and the MAE magnitudes are positively correlated with the temperature change from 20 $^\circ$C. The average MAE at 60 $^\circ$C goes above 1 (mS/cm), two orders of magnitudes higher than that in the interpolation case. Compared to the \textbf{GNN-only} baseline, we found a superior accuracy with \textbf{DiffMix}. The average MAE drops from 0.39 (mS/cm) to 0.24 (mS/cm), from 0.83 (mS/cm) to 0.49 (mS/cm), from 1.28 (mS/cm) to 0.75 (mS/cm), and from 1.74 (mS/cm) to 1.03 (mS/cm), at T of  30, 40, 50, 60 $^\circ$C, respectively. We further compare the \textbf{DiffMix} prediction results with the experimental measurements, as shown in the parity plot of \Cref{F3} (d). Both the interpolation and extrapolation testing results of \textbf{DiffMix} are validated by experimental measurements for the solutions of lithium hexafluorophosphate (\ce{LiPF6}) in ethylene carbonates (EC), propylene carbonates (PC), and ethyl methyl carbonates (EMC) solvent mixtures\cite{exp_ion}. In the experimentation, the salt concentration varied between 0.2 (mol/kg) and 2.1 (mol/kg), and the EC:PC ratio was varied with (EC+PC):EMC ratio fixed at 3:7 and 1:1, respectively. We find a good agreement between \textbf{DiffMix} predictions and experiments, even in the extrapolation test with temperatures higher than 20 $^\circ$C. Quantitatively, the $R^2$ and Pearson correlation coefficient values for interpolation and extrapolation sets are (0.80, 0.94) and (0.75, 0.92), and the interpolation and extrapolation MAEs are 0.75 (mS/cm) and 1.07 (mS/cm), respectively. Based on the results in \Cref{F3} (d), we conclude that the AEM-generated data provide an accurate basis to learn the complex electrolyte patterns via \textbf{DiffMix} at the given conditions. %We also observe a high standard deviation for the experimental results reported in \cite{exp_ion}, which may cause the deviation of predictions from experimental measurements. 

\begin{figure}[hbt!]
\centering
\includegraphics[width=\linewidth]{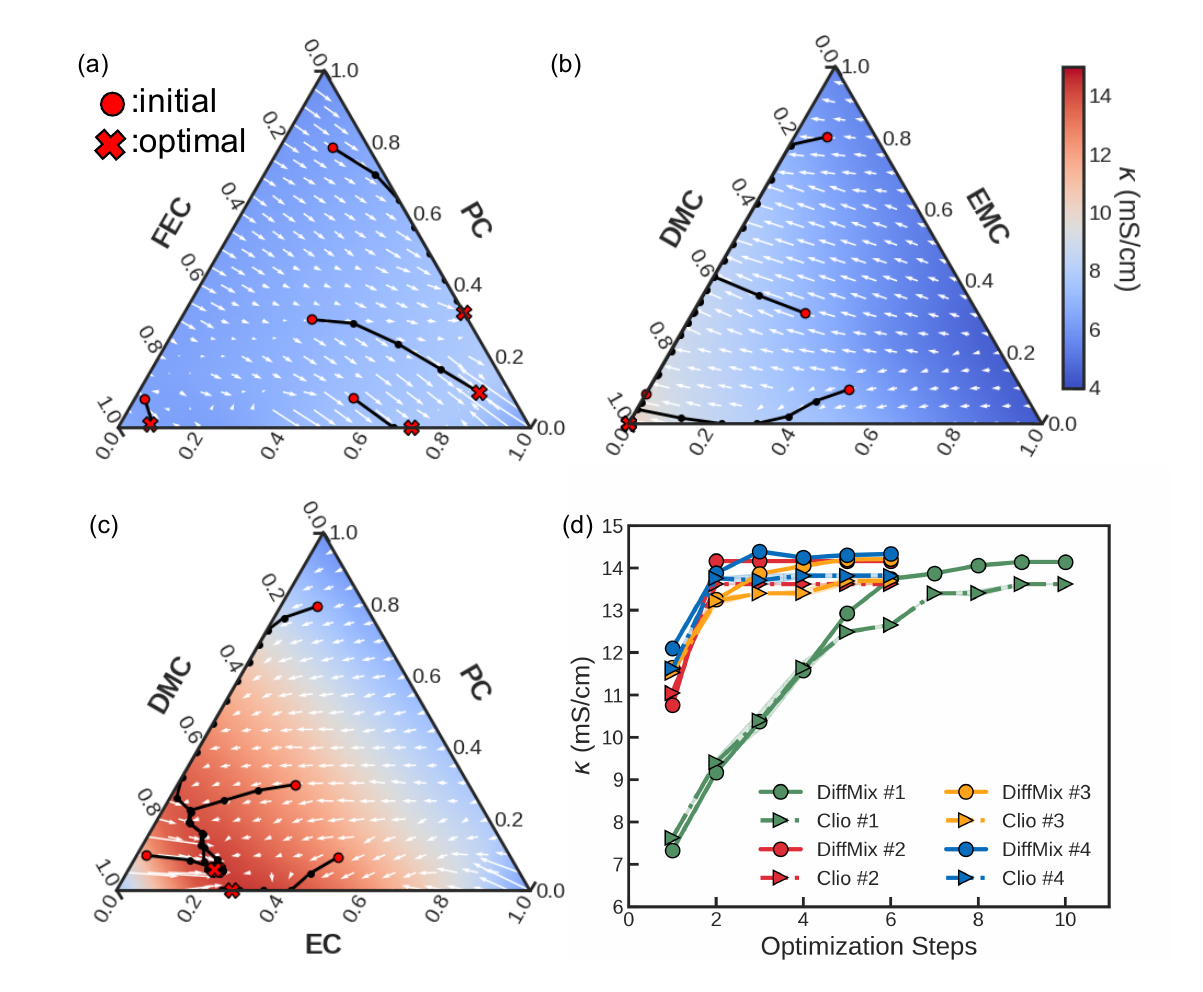}
\caption{\label{F4} Differentiable Battery Electrolyte Optimization with \textbf{DiffMix} and Robotic Experimentation. Optimizing on ionic conductivity ($\kappa$) landscape (Temperature=30 $^\circ$C) for (a) \ce{LiPF6} salts dissolved in PC:FEC:EC mixtures (with fixed lithium-ion mole fraction of 0.08); (b) \ce{LiPF6} salts dissolved in EMC:DMC:DEC mixtures (with fixed lithium-ion mole fraction of 0.12); (c) \ce{LiPF6} salts dissolved in PC:EC:DMC mixtures (with fixed lithium-ion mole fraction of 0.08). In each optimization case, a batch of four trajectories has been simulated starting from the dot sign and ending at the cross sign. The white arrows are the gradient information obtained by auto-differentiating \textbf{DiffMix}. (d) Optimization Curve of ionic conductivities in (c) along the four trajectories, where we include both \textbf{DiffMix} results and the robotic experimentation results generated by \textbf{Clio}.} %(e) Ionic conductivities ($\kappa$, red curve) and their gradients ($\Delta\kappa$, blue curve) with varying lithium concentrations and fixed solvent composition (DMC:EC, 0.7:0.3). (f) Visualization of the optimization trajectories of ionic conductivities fixed solvent compositions (four initial points in landscape of (b)), where we include the experimental validation of \textbf{Clio}.
\end{figure}

\subsection*{Differentiable Battery Electrolyte Optimization with \textbf{DiffMix} and Robotic Experimentation}

Fast charging of Li-ion batteries is impacted by electrolyte ionic conductivities, and electrolyte optimization can be challenging for battery design due to high experimentation costs\cite{clio}. With the trained \textbf{DiffMix} model, we test its capability to evaluate ionic conductivities and design electrolyte mixtures for high-performing Li-ion batteries. We select three types of electrolyte solutions as test cases and evaluate their ionic conductivities at 30 $^\circ$C and varying co-solvent compositions. They are \ce{LiPF6} salt in solvent mixtures of (I) cyclic carbonates, including ethylene carbonates (EC), propylene carbonates (PC) and fluorinated ethylene carbonates (FEC), (ii) linear carbonates, including ethyl methyl carbonates (EMC), diethyl carbonates (DEC) and dimethyl carbonates (DMC) and (iii) cyclic and linear carbonates, including EC, PC, and DMC.
We first show the ionic conductivity landscape of (i) in \Cref{F4} (a) by varying co-solvent compositions with fixed lithium mole fractions of 0.08, where we observe a moderate ionic conductivity peak up to 8 (mS/cm) in the EC-enriched region. Note that we treat the anions and cations separately when computing the mole fraction. \Cref{F4} (b) provides the conductivity landscape of electrolyte mixture (ii), where the highest $\kappa$ values are observed in the DMC-enriched region. Here, we fixed the lithium mole fraction at 0.12 due to the low dielectric constants of linear carbonates and thus the low dissociation degree of lithium salts. According to the conductivity map of the electrolyte mixture (iii) shown in \Cref{F4} (c), adding linear carbonate molecules into cyclic carbonate solvents can significantly increase the mixture ionic conductivities, where the maximum ionic conductivity is 14.39 (mS/cm) when PC:DMC:EC ratio is close to 0:0.70:0.30 with a fixed lithium-ion mole fraction of 0.08. We verify this result with the output of our data generator, AEM\cite{aem}, which provides the highest conductivity of 14.2 (mS/cm) at 0.082 to 0.085 lithium mole fraction with the given PC:DMC:EC ratio. This agrees well with the differentiable modeling result. It is worth noting that the training data is produced with a temperature lower than 20 $^\circ$C, but we see a good generalization at 30 $^\circ$C.

As previously introduced, the gradient information is readily accessible by differentiating the trained \textbf{DiffMix} model. To illustrate that, we show the gradient vectors as arrows in the ionic conductivity landscapes in \Cref{F4} (a)-(c). In \Cref{F4} (a) and (c), we observe large gradients at pure EC solvent area, indicating that adding a small number of co-solvents can significantly improve the ionic conductivity. This can be explained by EC's being solid-like at room temperature\cite{le3}. Another interesting observation is that pure DMC solvent area in \Cref{F4} (c) displays a much higher gradient than that in \Cref{F4} (b), with which we conclude that adding a small quantity of high-polarity cyclic carbonate solvents (EC, PC) could enable higher ionic conductivities. Based on the gradient information provided by \textbf{DiffMix}, we implement a gradient-ascent algorithm by initializing a batch of four starting points in the mixture space and increasing the objectives iteratively following the gradient directions. From \Cref{F4} (a)-(c), our optimization algorithm robustly identified local maximum spots. %and meanwhile reduced the number of evaluations, accelerating  the discovery of optimal electrolyte formulations
This differentiable optimization framework further guides the robotic experimentation performed by our hardware setup, \textbf{Clio}. We extract the batch of four optimization trajectories in \Cref{F4} (c) and compare the ionic conductivities evaluated at each step by both \textbf{DiffMix} and \textbf{Clio}, as shown in \Cref{F4} (d). \textbf{DiffMix} and \textbf{Clio} results show a good agreement between simulation and experimentation, and we show at least an 18.8\% increase from the initial ionic conductivities. %This further elucidates the power of our differentiable mixture physics model. 
%Across \Cref{F4} (a)-(d), we focus on varying solvent composition space while fixing lithium mole fractions. \Cref{F4} (e) and (f) further introduces the optimization scenario of fixed solvent compositions and varying salt concentrations. \Cref{F4} (e) is created given the DMC:EC ratio of 0.7:0.3, close to the bottom ending point of \Cref{F4} (b), indicating that the optimal lithium mole fraction is around 0.08, agreeing with the value we select in \Cref{F4} (b). \textbf{Clio} further validated the \textbf{DiffMix} simulation results in \Cref{F4} (f). 
These results elucidate the capability of differentiable modeling of battery electrolytes, with which we could efficiently explore the chemical space of multi-component electrolyte mixtures.% considering the significant complexity of mixtures on its composition space in addition to the ultra-large component space of over billions of molecules\cite{doi:10.1021/ci300415d}.

\section*{Discussion}

% In this work, focusing on battery electrolytes, we develop a differentiable modeling and optimization framework for chemical mixtures, \textbf{DiffMix}, that combines the advantages of physics-based models and geometric deep learning, and further guides the robotic experimentation for the practical electrolyte optimization.
In this work, focusing on battery electrolytes, we develop a GDL-based differentiable model for chemical mixtures, \textbf{DiffMix}, that combines the advantages of physics-based models and geometric deep learning and further guides the robotic experimentation for practical electrolyte optimization.
%to learn the mixture representations and predict their properties with high accuracy and efficiency. We start by introducing polynomial-based thermodynamic laws and transport laws for chemical mixtures. We replace chemistry-dependent physical constants by GDL learnable functions and enable the end-to-end differentiability from molecule structures to macroscopic mixture properties. 
The evaluation results on thermodynamic data of binary non-electrolyte mixtures and ion transport data of electrolyte mixtures indicate that \textbf{DiffMix} preserves the component-wise permutation invariance and enables more accurate and robust predictions than \textbf{GNN-only} and \textbf{MixECFP} (in Table S 1), as can be seen from the low MAEs and MAE variances. When extrapolated to high temperatures, \textbf{DiffMix} predictions show superior accuracy than the \textbf{GNN-only} baseline, due to the incorporation of a temperature-dependent VFT model. The experimental measurements and \textbf{DiffMix} display a good agreement with each other, even in the extrapolation case, enabling the real-world applications of our trained model.

We further test the physics model capacity of R-K thermodynamic mixing law in \textbf{DiffMix} by tuning the polynomial order $N$ in \Cref{eq3} and observe a plateauing behavior beyond $N=4$. A distinction between excess molar volumes and enthalpies is observed that the linear mixing model outperforms the \textbf{GNN-only} model only for the volume task but not for the enthalpy task, which can be explained by the high inter- and intra-molecular interaction dependence of enthalpies. Although this demonstrates the flexibility of our model in terms of the function forms of physics laws, future investigation is required to explore other types of thermodynamic and kinetic laws for mixtures.

By building our model in a fully differentiable framework, gradient information is readily accessible for a trained \textbf{DiffMix} model. This further allows us to optimize ionic conductivity over the input space. Taking the input co-solvent composition as variables, we identify peak ionic conductivity areas for various ternary co-solvent electrolyte chemical spaces by mixing linear carbonate solvents and cyclic carbonate solvents. The simulated trajectories have been utilized to guide the robotic experimentation performed by \textbf{Clio}, which successfully increases ionic conductivity values by over 18.8\%. It is worth noting that in this work we conduct the DiffMix-guided robotic experimentation in a two-step process, (1) training a \textbf{DiffMix} model with simulated AEM data and running the optimization on the modeled response surface, (2) guiding \textbf{Clio} with the predefined optimization trajectory. In an alternative way, especially when the simulation is not of high quality, a closed-loop optimization can be designed via retraining the \textbf{DiffMix} model every few iterations during experimental data collection, which may enable a more robust and adaptive optimization. Our work has expanded the modeling and optimization techniques of battery electrolyte mixtures by unifying physics laws and geometric deep-learning in a differentiable framework.
%In the future, it is worth exploring other types of mixture physics laws, or automating the mixing law identification by symbolic regression\cite{sr1,sr2}.% Other future work may entail exploring more complex multi-solvent dual-salt electrolytes\cite{dual} or high-entropy electrolytes\cite{hee1,hee2}.

\section*{Methods}

\subsection*{Data Collection and Generation}
Thermodynamic and transport mixture property datasets were prepared for benchmarking models developed in this work. The thermodynamic datasets include excess molar enthalpy \cite{ee1,ee2,emv6,ee4} and excess molar volume\cite{emv1, emv2, emv3,emv4,emv5,emv6,emv7,emv8} values curated from the literature. There are 631 data points for excess molar enthalpy, covering 34 unique mixture chemistries with varying compositions. For excess molar volume, there are 1,069 binary mixture data points based on 28 unique mixtures composed of 25 organic chemicals.  For ionic conductivities, we prepared an ionic conductivity dataset that contains over 24,822 mixtures of single-salt-ternary-solvent electrolyte solutions. These electrolyte components consist of two unique salt species, including lithium hexafluorophosphate (\ce{LiPF6}), lithium bis((trifluoromethyl)sulfonyl)azanide (LiTFSI), and six organic carbonate solvents, including ethylene carbonates (EC), propylene carbonates (PC), fluorinated ethylene carbonates (FEC), ethyl methyl carbonates (EMC), diethyl carbonates (DEC) and dimethyl carbonates (DMC). The electrolyte data was generated with one salt and any arbitrary combinations of three co-solvents, with the salt concentration ranged in $\{0.025, 0.5, 1.0, 1.5, 2.0, 2.5, 3.0\}$ molal and each-solvent mass fractions varying from $\{0,0.2,0.4,0.6,0.8,1.0\}$. The data generation was performed by the Advanced electrolyte model (AEM) that produces high-fidelity electrolyte data for the chemical species evaluated here\cite{aem}. In the collected datasets,  molecular identities were stored in their Simplified Molecular Input Line Entry System (SMILES) format\cite{smiles}, from which we can retrieve the topology and chemical information on atoms and bonds with RDKit\cite{rdkit}. We converted all compositions into mole fractions for model training in the next section.

\subsection*{Model Implementation and Training}

For \textbf{GNN-only} and \textbf{DiffMix}, the atom features considered include one hot encoding of atom type, number of heavy neighbors, formal charges, hybridization type, chirality, and number of implicit hydrogens, and numerical information on ring structures, aromaticity, atomic mass, VdW radius, and covalent radius, giving a 97-dimension feature vector. Note that no bond features are incorporated in our model, but can be included in future work. The graph convolution cell is made up of 3 \textbf{GraphConv} \cite{graphconv} steps, each of which is followed by ReLU and dropout layers (dropout rate, p=0.25). The whole graph convolution cell ends up with a global mean pooling layer and provides graph-level embeddings for component molecules $g'_i$. These graph-level embeddings are then concatenated with compositions and environment conditions. 

For \textbf{GNN-only}, no physics laws are incorporated, and therefore %the substance embeddings are transformed by a feed-forward neural network \textbf{SubNet}. To further improve the model capacity, we add one more neural network to learn the pairwise embeddings by transforming pairs of substance embeddings, which is called \textbf{PairNet}. The transformed substance and pair embeddings are weighed averaged to obtain the mixture embeddings, which are processed by the last feed-forward neural network, \textbf{MixtureNet}. The 
\textbf{MixtureNet} output is the predicted mixture property. The dimensions of \textbf{SubNet}, \textbf{PairNet}, and \textbf{MixNet} go as follows: $[N,N,N,N], [2N,2N,2N,N], [2N,4N,2N,1]$, where $N=256+1+Nenv$ and $Nenv$ is the dimension of environment conditions. For \textbf{DiffMix}, it is treated differently for thermodynamic and ion transport properties, since distinct mixing laws are selected. With the VFT model selected, \textbf{MixNet} now is changed into $[2N,4N,2N,3]$, which outputs the three physical coefficients in the VFT model.  In terms of the thermodynamics of mixing, \textbf{SubNet} and \textbf{PairNet} are reused to obtain the component-wise physical parameters $\{P^i\}$ and pair-wise physical parameters $\{C^k_{RK,ij}\}$. Instead of creating a \textbf{PairNet} for each order k, we designed it as a multitask neural network, which shares parameters before the output layers. 

During training, we set the learning rate as 0.001 with a weight decay rate of $10^{-4}$ in PyTorch\cite{pytorch} Adam optimizer. L1Loss is used for loss backpropagation. We also applied early stopping criteria to select the epoch with the lowest validation error to avoid overfitting. The ionic conductivity labels went through a logarithm transformation before computing the loss values to distinguish mixture properties that span multiple orders of magnitudes. 
All models were implemented with PyTorch\cite{pytorch} and PyTorch Geometric\cite{pyg}. For each mixture property, the full dataset was randomly split into training, validation, and testing sets, in the ratio of $8:1:1$. The cross-validation results were reported based on an ensemble of 5 models with randomly varying splits. %The full package and data will be coming soon at \href{https://github.com/BattModels/DiffMix.git}{https://github.com/BattModels/DiffMix.git}. 

% \subsection*{Gradient-based Optimization with DiffMix and Clio} Differentiable physics models enable gradient-based optimization for materials modeling and design\cite{GUAN2022114217}. One powerful capability of \textbf{DiffMix} is the auto-differentiation from mixture property to input chemistry or compositions, which provides gradient information at any arbitrary compositions. We build a gradient-ascent optimization algorithm to identify peak ionic conductivities of battery electrolytes. The input variables are compositions in this case, but can be expanded to molecule species in a future study. Furthermore, we integrate our optimization trajectories into an automated experimentation setup, \textbf{Clio}\cite{clio}, and run \textbf{DiffMix}-derived gradient-ascent optimization trajectories.

\subsection*{Automated Experimental Measurements of Electrolyte Properties}
The ionic conductivity measurements in this work were done by Clio, a custom-built robotic setup, developed previously in our group \cite{clio}. The ionic conductivity data were measured by electrochemical impedance spectroscopy (EIS) in a PTFE fixture chamber using a PalmSens4 impedance analyzer. The electrolytes were filled into the chamber between two symmetric Pt electrodes. The impedance of the cell was measured at five frequencies between 14 kHz and 800 kHz. The resistance of the sample is determined by evaluating the real part of the impedance at the frequency where the smallest phase difference is observed during measurement. To calculate the specific ionic conductivity of the sample, a cell constant is obtained through a single point calibration using a known solution (Acetonitrile and \ce{LiPF6}). The specific ionic conductivity is then determined by dividing the inverse resistance by the cell constant. The temperature was managed via glove-box-wide heating and airflow. Temperatures were $27.2^{\circ}$C $\pm 0.3^{\circ}$C. We note that this is slightly lower than the predictions of \textbf{DiffMix}, thus temperature may account for deviations between experimental and modeled data. %The measurement temperature values $(\sim27^{\circ}C)$ are slightly below simulation temperature$(30^{\circ}C)$.

\subsection*{Experimental Methods: Materials Availability}
The electrolyte salt (LiPF6) and solvents (PC, DMC, EC) used in this study were obtained from Linyi Gelon LIB Co. Ltd., anhydrous ($<$20 ppm) and battery grade (99.9\% pure). The precursors and electrolyte stock solutions were prepared and stored in a dry Ar-filled glove box ($<$100 ppm oxygen, $<$10 ppm H2O).
The stock solutions were made by first mixing the solvents into the desired mass ratios, then gradually adding salts to the solvents to the designated concentrations. The mass of the solutes and solvents were measured using a Denver Instrument PI-214.1 analytical balance. All solutions were mixed with a magnetic stir bar and magnetic stir plate in a glass beaker for at least half an hour after the last visible salt. The solutions were then transferred to and stored in 60-mL amber glass vials with Sure/Seal septa lids.

\subsection*{Data Availability}
Our data and code will be coming soon at \href{https://github.com/BattModels/DiffMix.git}{https://github.com/BattModels/DiffMix.git}. % More details on data processing and experimental materials preparation can be found in SI. 

\section*{Author Contributions}
S.Z. and V.V. designed research; S.Z., B.R. E.A., P.-W.G., and V.V., contributed to the conceptualization and methodology of the DiffMix framework; S.Z. implemented the algorithms; H.L. and A.D. designed and performed the experiments; All authors analyzed data and wrote the paper.

\section*{Competing Interest Statement}

V.V., S.Z., and B.R. are inventors on a patent application related to predicting and optimizing mixture properties by geometric deep learning. P.-W.G. and V.V. are inventors on a patent application related to system and method for material modelling and design using differentiable models.

\begin{acknowledgement}

We acknowledge funding from the Advanced Research Projects Agency-Energy (ARPA-E), U.S. Department of Energy, under Award Number DE-AR0001211. The views and opinions of authors expressed herein do not necessarily state or reflect those of the United States Government or any agency thereof. H. L., A. D., and V. V. acknowledge the support of Toyota Research Institute through the Accelerated Materials Design and Discovery program. S. Z. and V. V. acknowledge support from the Extreme Science and Engineering Discovery Environment (XSEDE) for providing computational resources, under Award Number TG-CTS180061. We also acknowledge Dr. Jay Whitacre, Dr. Andrew Li, Dr. Lei Zhang, and others in Venkat's group for their support and discussions.
\end{acknowledgement}

%%%%%%%%%%%%%%%%%%%%%%%%%%%%%%%%%%%%%%%%%%%%%%%%%%%%%%%%%%%%%%%%%%%%%
%% The same is true for Supporting Information, which should use the
%% suppinfo environment.
%%%%%%%%%%%%%%%%%%%%%%%%%%%%%%%%%%%%%%%%%%%%%%%%%%%%%%%%%%%%%%%%%%%%%
% \begin{suppinfo}

% In the Supporting Information, we provide mathematical details of $\texttt{O-GNN}$, data distribution and supporting figures. Code is available at \href{https://github.com/shangzhu-cmu/GNN_EST}{https://github.com/shangzhu-cmu/GNN\_EST}. 

% \end{suppinfo}

%%%%%%%%%%%%%%%%%%%%%%%%%%%%%%%%%%%%%%%%%%%%%%%%%%%%%%%%%%%%%%%%%%%%%
%% The appropriate \bibliography command should be placed here.
%% Notice that the class file automatically sets \bibliographystyle
%% and also names the section correctly.
%%%%%%%%%%%%%%%%%%%%%%%%%%%%%%%%%%%%%%%%%%%%%%%%%%%%%%%%%%%%%%%%%%%%%
\bibliography{achemso-demo}


\section*{Supplementary material}

# Supplementary Information for "Differentiable Modeling and Optimization of Battery Electrolyte Mixtures Using Geometric Deep Learning"

Shang Zhu,[†] Bharath Ramsundar,[‡] Emil Annevelink,[†] Hongyi Lin,[†] Adarsh Dave,[†] Pin-Wen Guan[a],[†] Kevin Gering,[¶] and Venkatasubramanian Viswanathan[*,†]

[†]Department of Mechanical Engineering, Carnegie Mellon University, Pittsburgh, Pennsylvania 15213, USA

[‡]Deep Forest Sciences, Palo Alto, California 94306, USA

[¶]Energy Storage & Technology, Idaho National Laboratory, Idaho Falls, Idaho 83415, USA

E-mail: venkvis@cmu.edu

---

[a]Current affiliation: Sandia National Laboratories, Livermore, California 94550, USA

# S1: Baseline Model with Morgan Fingerprints

We further design a purely data-driven baseline model, **MixECFP**, by removing the mixture physics model and replacing the GNNs with Morgan fingerprints.[1] **MixECFP** simply concatenates the fingerprints, compositions, and environment conditions, so it does not include a mixture pooling operator and does not preserve the permutation invariance. Overall, the results of three model variants are shown in Table S 1, **DiffMix**, **GNN-only** and **Mix-**

1

**ECFP**. In **MixEFCP**, by permuting the sequences of mixture components, the testing MAE increases significantly for all three tasks, preventing its use in real-world applications. Further, **DiffMix** and **GNN-only** outperform **MixECFP** on the model accuracy even without permuting.

Table S 1: Model Performance on Thermodynamic Properties and Ionic Conductivities[a]

| Task | DiffMix test | GNN-only test | MixECFP test | MixECFP permuted[c] |
|---|---|---|---|---|
| $V_m^E$ (cm$^3$/mol) | **0.033±0.009**[b] | 0.090±0.106 | 0.034±0.008 | 0.462±0.114 |
| $H_m^E$ (J/mol) | **5.10±0.32**[b] | 9.88±2.21 | 16.97 ± 2.19 | 365.26 ± 44.53 |
| $\kappa$ ($mS/cm$) | **0.044±0.005** | 0.045±0.006 | 0.105 ± 0.007 | 22.70 ± 6.62 |

*a*. Results are reported by regression mean-absolute-errors (MAEs) (Mean ± Standard Deviation) after running an ensemble of 5 models. *b*. The polynomial order $N$ is 4 in Equation (2) in Main Text. *c*. Permuted loss is generated by permuting input component sequences.

## S2: Data Efficiency of DiffMix

The amount and diversity of the data may highly influence the performance of data-driven models. To evaluate the model performances on the datasets of varying sizes, we randomly sampled 10%, 20%, 50%, and 100% of the original thermodynamic datasets and reran the training processes in order to obtain a curve of testing losses against the amount of sampled data. The curve is visualized in Figure S 1 (a) and (b), indicating the deteriorated model accuracy when fewer data were provided. From Figure S 1 (a) and (b), we found that **DiffMix** is more accurate than the **GNN-only** baseline, except for the two most data-limited points in Figure S 1 (a) on excess molar volumes. This may be attributed to the high variances of the model performances.

## S3: Physics Capacity Analysis on DiffMix

To further interpret the trained **DiffMix** model on thermodynamic data in this work, we visualize the magnitudes of R-K coefficients $\{C_{RK,ij}^k\}$ varied by the polynomial order number

2

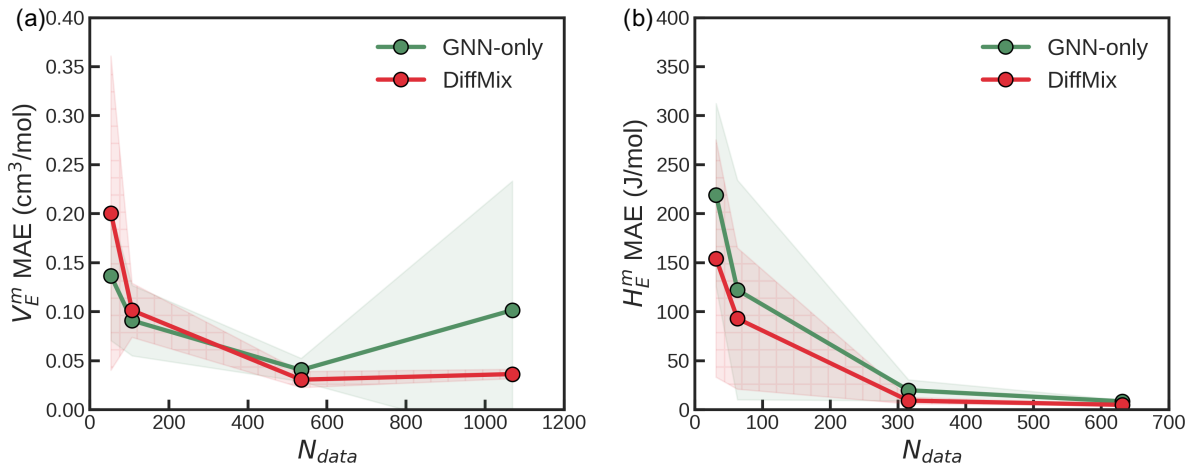

Figure S 1: (a-b) The testing regression MAEs as a function of the number of training data for (a) excess molar volumes ($V_m^E$), and (b) excess molar enthalpies ($H_m^E$). These curves are generated after running an ensemble of 5 models, where the solid lines and shaded areas display the mean values standard deviations of results, respectively.

$k$, as shown in Figure S 2. Both the results in Figure S 2 (a) and (b) indicate an overall decreasing trend of the R-K coefficients with some fluctuations, where the zeroth order ($k$=0) term plays the most dominant role in the following R-K polynomials.

$$P_m = \Sigma_{i<j}[x_i x_j \Sigma_{k=0}^{N} C_{RK,ij}^k (x_i - x_j)^k] + \Sigma_i x_i P^i \quad (1)$$

This is consistent with the plateauing pattern of the model performance when increasing the number of polynomials involved (N), where adding higher-order terms does not help improve the model accuracy, as discussed in the main text.

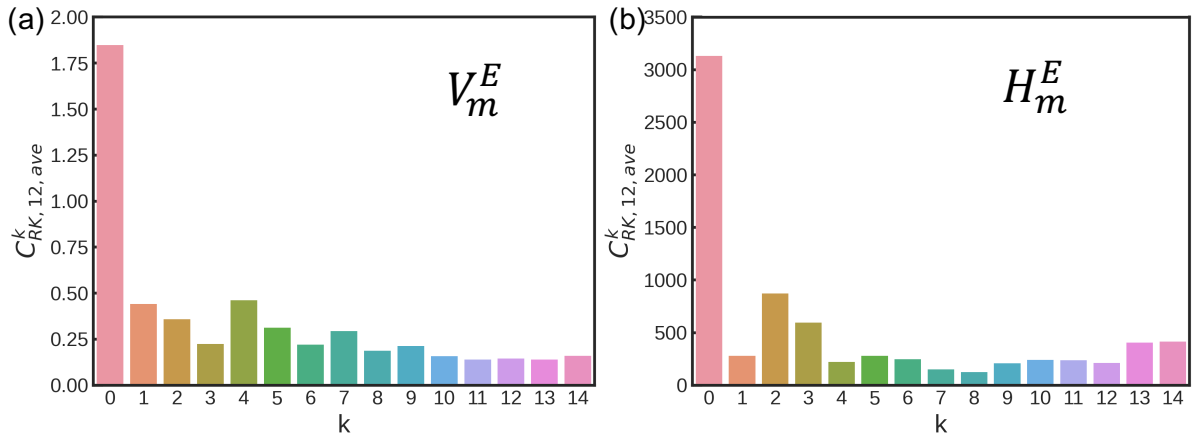

Figure S 2: Magnitudes of R-K Coefficients $\{C^k_{RK,ij}\}$ Varied by the Order Number $k$, on the Task of (a) Excess Molar Volumes ($V^E_m$), and (b) Excess Molar Enthalpies ($H^E_m$). The coefficients are extracted by averaging the values from all data points, including training, validation and testing, as well as from all five models in the ensemble of the model training process (R-K order numbers of these models (N) are set as 15 during training). For the binary systems evaluated in this work, only one series of interaction coefficients $\{C^k_{RK,12}\}$ exists, i.e. between chemical species 1 and chemical species 2.

# S4: Differentiable Optimization Results with Fixed Solvent Concentrations

Across Figure 4 (a)-(d) in the main text, we focus on varying solvent composition space while fixing lithium mole fractions. Figure S 3 (a) and (b) further introduce the optimization scenario of fixed solvent compositions and varying salt concentrations. Figure S 3 (a) is created given the DMC:EC ratio of 0.7:0.3, close to the bottom ending point of Figure 4 (c) in the main text, indicating that the optimal lithium mole fraction is around 0.08, agreeing with the value we select in Figure 4 (c) in the main text. **Clio**[2] further validated the **DiffMix** simulation results in Figure S 3 (b).

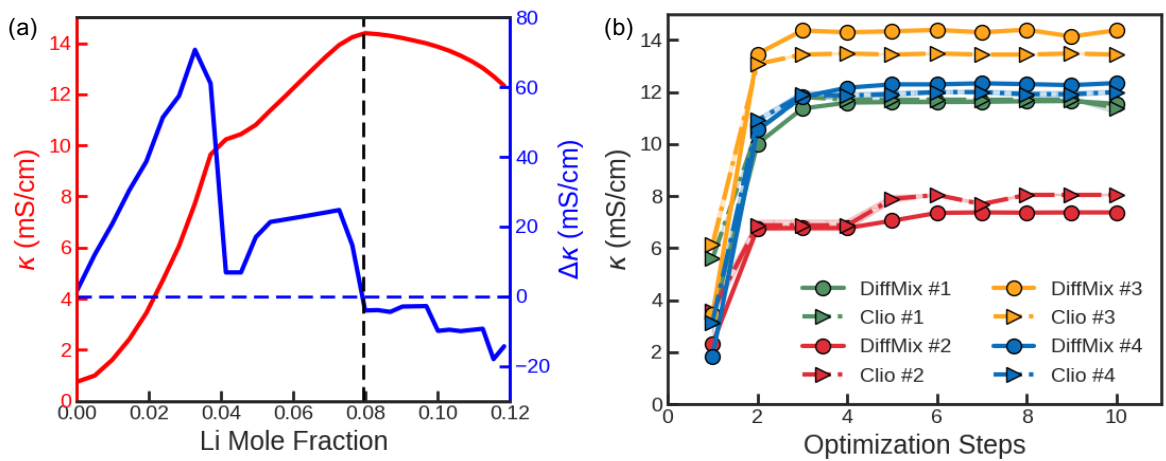

Figure S 3: (a) Ionic conductivities ($\kappa$, red curve) and their gradients ($\Delta\kappa$, blue curve) with varying lithium concentrations and fixed solvent composition (DMC:EC, 0.7:0.3 mole fraction). (b) Optimization Curve of ionic conductivities with fixed solvent compositions (four initial points Figure 4 (c) in the Main Text), where we include both **DiffMix** results and the robotic experimentation results generated by **Clio**.

## Differentiable Optimization Data

In Table S 2, we present the data generated from differentiable optimization and compare the results from both **DiffMix** and **Clio**. These data correspond to Figure 4 (c-d) in the main text. Note that the lithium-ion mole fraction is fixed as 0.08 and the temperature we run **DiffMix** optimization is 30 °C, while the temperature of robotic experimentation is slightly off.

Table S 2: Differentiable Optimization History[a]

| batch | step | $PC_{mol}$ | $DMC_{mol}$ | $EC_{mol}$ | $\kappa_{ML}$ (mS/cm) | $\kappa_{Clio}$ (mS/cm) | $T_{Clio}$ (°C) |
|---|---|---|---|---|---|---|---|
| 1 | 0 | 0.8 | 0.1 | 0.1 | 7.33 | 7.63±0.02 | 27.16±0 |
| 1 | 1 | 0.73 | 0.26 | 0 | 9.17 | 9.41±0.09 | 27.06±0.02 |
| 1 | 2 | 0.65 | 0.35 | 0 | 10.37 | 10.39±0.22 | 27±0.01 |
| 1 | 3 | 0.54 | 0.46 | 0 | 11.58 | 11.65±0.16 | 26.94±0 |
| 1 | 4 | 0.4 | 0.6 | 0 | 12.93 | 12.49±0.01 | 26.88±0.01 |
| 1 | 5 | 0.27 | 0.71 | 0.03 | 13.74 | 12.65±0.05 | 26.94±0.03 |
| 1 | 6 | 0.19 | 0.71 | 0.1 | 13.87 | 13.4±0 | 27.06±0 |
| 1 | 7 | 0.13 | 0.71 | 0.15 | 14.06 | 13.41±0.08 | 27.13±0.02 |
| 1 | 8 | 0.06 | 0.75 | 0.19 | 14.14 | 13.62±0.01 | 26.99±0.01 |
| 1 | 9 | 0.06 | 0.75 | 0.19 | 14.14 | 13.62±0.01 | 26.99±0.01 |
| 2 | 0 | 0.1 | 0.9 | 0 | 10.76 | 11.05±0 | 26.89±0 |
| 2 | 1 | 0.06 | 0.74 | 0.19 | 14.16 | 13.62±0.01 | 26.99±0.01 |
| 2 | 2 | 0.06 | 0.74 | 0.2 | 14.16 | 13.62±0.01 | 26.99±0.01 |
| 2 | 3 | 0.06 | 0.74 | 0.2 | 14.16 | 13.62±0.01 | 26.99±0.01 |
| 2 | 4 | 0.06 | 0.75 | 0.2 | 14.16 | 13.62±0.01 | 26.99±0.01 |
| 2 | 5 | 0.06 | 0.75 | 0.2 | 14.16 | 13.62±0.01 | 26.99±0.01 |
| 3 | 0 | 0.3 | 0.4 | 0.3 | 11.65 | 11.52±0.05 | 27.42±0.03 |
| 3 | 1 | 0.26 | 0.59 | 0.15 | 13.25 | 13.22±0.12 | 27.53±0.02 |
| 3 | 2 | 0.2 | 0.71 | 0.09 | 13.86 | 13.4±0 | 27.06±0 |
| 3 | 3 | 0.14 | 0.71 | 0.15 | 14.05 | 13.41±0.08 | 27.13±0.02 |
| 3 | 4 | 0.07 | 0.71 | 0.22 | 14.2 | 13.69±0.08 | 27.2±0.01 |
| 3 | 5 | 0.06 | 0.71 | 0.23 | 14.21 | 13.69±0.08 | 27.2±0.01 |
| 4 | 0 | 0.1 | 0.4 | 0.5 | 12.1 | 11.62±0.08 | 26.86±0.02 |
| 4 | 1 | 0 | 0.57 | 0.43 | 13.88 | 13.75±0.02 | 27.01±0.04 |
| 4 | 2 | 0 | 0.7 | 0.3 | **14.39**[b] | 13.71±0.14 | 27.2±0.03 |
| 4 | 3 | 0 | 0.74 | 0.26 | 14.24 | 13.82±0.04 | 27.2±0.02 |
| 4 | 4 | 0 | 0.73 | 0.27 | 14.3 | 13.82±0.04 | 27.2±0.02 |
| 4 | 5 | 0 | 0.73 | 0.27 | 14.33 | 13.82±0.04 | 27.2±0.02 |

*a.* Results are reported for four trajectories corresponding to Figure 4 (c-d) in the main text. The lithium mole fraction is fixed at 0.08. $\kappa_{ML}$ is **DiffMix** output with temperature set at 30 °C. *b.* Highest $\kappa_{ML}$ from **DiffMix** optimization, with **AEM**[3] verification of 14.2 (mS/cm) at the optimal lithium mole fraction of 0.082 to 0.085 with the given PC:DMC:EC ratio.


\end{document}